\documentclass[aps, prx, floats, twocolumn ,showkeys , superscriptaddress, floatfix, longbibliography]{revtex4-2}
\setcitestyle{super}
\usepackage{multirow}
\usepackage{xcolor}
\usepackage{amsmath}
\usepackage{graphicx} 
\usepackage[colorlinks=true, linkcolor = blue, urlcolor  = blue, citecolor = blue, anchorcolor = blue]{hyperref}
\usepackage{upgreek}

\makeatletter
\renewcommand{\p@subsection}{}
\renewcommand{\p@subsubsection}{}
\makeatother
\begin{document}
\title{\large Volume electron microscopy in injured rat brain validates white matter microstructure metrics from diffusion MRI}
\author{Ricardo Coronado-Leija}
\thanks{ricardo.coronadoleija@nyulangone.org, rleija@cimat.mx}
\affiliation{Bernard and Irene Schwartz Center for Biomedical Imaging, Department of Radiology, New York University School of Medicine, New York, NY, USA}
\author{Ali Abdollahzadeh}
\affiliation{Bernard and Irene Schwartz Center for Biomedical Imaging, Department of Radiology, New York University School of Medicine, New York, NY, USA}
\author{Hong-Hsi Lee}
\affiliation{Athinoula A. Martinos Center for Biomedical Imaging, Department of Radiology, Massachusetts General Hospital, Harvard Medical School, Boston, MA, USA}
\author{Santiago Coelho}
\affiliation{Bernard and Irene Schwartz Center for Biomedical Imaging, Department of Radiology, New York University School of Medicine, New York, NY, USA}
\author{Benjamin Ades-Aron}
\affiliation{Bernard and Irene Schwartz Center for Biomedical Imaging, Department of Radiology, New York University School of Medicine, New York, NY, USA}
\author{Ying Liao}
\affiliation{Bernard and Irene Schwartz Center for Biomedical Imaging, Department of Radiology, New York University School of Medicine, New York, NY, USA}
\author{Raimo A. Salo}
\affiliation{A.I. Virtanen Institute for Molecular Sciences, University of Eastern Finland, Kuopio, Finland}
\author{Jussi Tohka}
\affiliation{A.I. Virtanen Institute for Molecular Sciences, University of Eastern Finland, Kuopio, Finland}
\author{Alejandra Sierra}
\affiliation{A.I. Virtanen Institute for Molecular Sciences, University of Eastern Finland, Kuopio, Finland}
\author{Dmitry S. Novikov}
\affiliation{Bernard and Irene Schwartz Center for Biomedical Imaging, Department of Radiology, New York University School of Medicine, New York, NY, USA}
\author{Els Fieremans}
\affiliation{Bernard and Irene Schwartz Center for Biomedical Imaging, Department of Radiology, New York University School of Medicine, New York, NY, USA}
\begin{abstract} 
\noindent
Biophysical modeling of diffusion MRI (dMRI) offers the exciting potential of bridging the gap between the macroscopic MRI resolution and microscopic cellular features, effectively turning the MRI scanner into a noninvasive \textit{in vivo} microscope. In brain white matter, the Standard Model (SM) interprets the dMRI signal in terms of axon dispersion, intra- and extra-axonal water fractions and diffusivities. However, for SM to be fully applicable and correctly interpreted, it needs to be carefully evaluated using histology. Here, we perform a comprehensive histological validation of the SM parameters, by characterizing WM microstructure in sham and injured rat brains using volume (3d) electron microscopy (EM) and \textit{ex vivo} dMRI. Sensitivity is evaluated by how close each SM metric is to its histological counterpart, and specificity by how independent it is from other, non-corresponding histological features. This comparison reveals that SM is sensitive and specific to microscopic properties, clearing the way for the clinical adoption of \textit{in vivo} dMRI derived SM parameters as biomarkers for neurological disorders. 
\end{abstract}
\maketitle
\section{INTRODUCTION}
Diffusion MRI (dMRI) has the remarkable ability \citep{Kiselev2017,Novikov2018b,Alexander2017}
to probe the Brownian motion of water molecules on length scales comparable to cellular structures ($\sim$ 10 $\upmu$m), two orders of magnitude smaller than the standard clinical MRI resolution ($\sim$ 1 mm). In this way, dMRI carries information about the cellular-level architecture restricting the diffusion of water molecules, and offers the exciting potential to characterize tissue microstructure \textit{in vivo} and non-invasively in clinical settings \citep{Jelescu2017}. However, dMRI signals are indirect measures of tissue structural properties, and the task of identifying and extracting the relevant information remains an active field of research \citep{Novikov2018b,Jelescu2017,Alexander2017,Novikov2018a,Weiskopf2021}.

Biophysical modeling of the dMRI signal strives to provide metrics that are not only sensitive but also specific to the underlying tissue microstructure \citep{Jelescu2017,Novikov2018a}. Biophysical models rely on assumptions (they ``draw a picture") about the geometry of the tissue that restricts the diffusion of the water. If these assumptions accurately capture the relevant properties that are encoded in the dMRI signal, such biophysical models can bridge the gap between the millimeter macroscopic MRI voxel resolution and the micrometer-level tissue architecture.

\begin{figure*}[htb!]
\centering
\includegraphics[width=\textwidth]{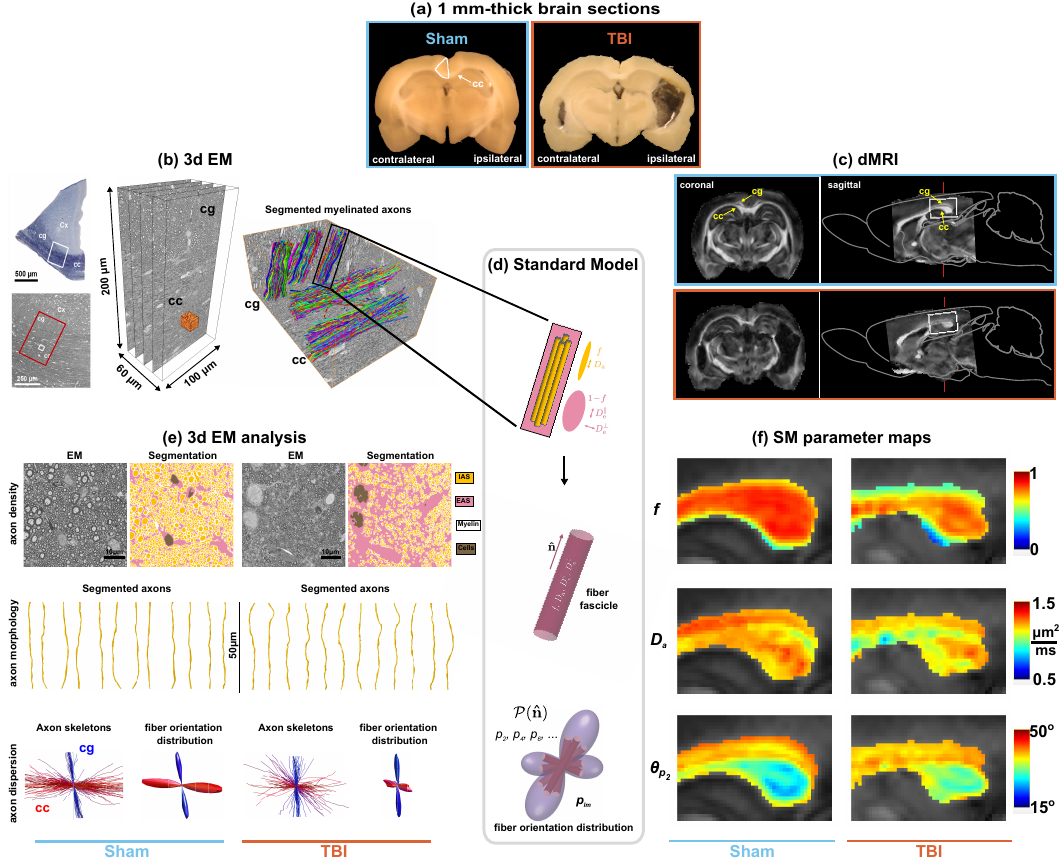}
\caption{\textbf{SM parameter maps reveal specific microstructure properties that correspond to 3d EM histology}. (a) Brain sections (1 mm-thick) of sham-operated and TBI rats. (b) Location of the 3d EM samples within sham-operated rat brain and automatically segmented myelinated axons. (c) FA maps of rat brains in (a), indicating the corpus callosum (cc) and cingulum (cg) at the location where the 3d EM samples were collected, brain sagittal contours adapted from \citep{Swanson2018}. (d) In the Standard Model, a fiber fascicle is modeled by an intra-axonal compartment, in which axons are represented by cylinders with zero-radius, and an extra-axonal compartment represented by an axially symmetric tensor (see {\it Methods}). These compartments are described by parameters $f$, $D_\mathrm{a}$, $D_\mathrm{e}^{\parallel}$ and $D_\mathrm{e}^{\perp}$. One voxel contains multiple orientationally dispersed fascicles that contribute to the directional dMRI signal according to their FOD $\mathcal{P}(\hat{n})$. (e) EM images transverse to the axons of the cingulum for the rat brains in (a), with segmented IAS, EAS, and myelin. Representative segmented 3d axons are also shown. Using skeletons from thousands of segmented axon, FODs for each axon population in the volumes were constructed. Reduction in axon density, changes in axon morphology, and increase in axon dispersion can be observed for the TBI animal with respect to the sham-operated animal. (f) dMRI-derived SM parametric maps reveal specific differences between sham and TBI that are observed in corresponding 3d EM histology.}
\label{fig:overview}
\end{figure*}

In brain white matter (WM), at sufficiently long diffusion times, the dMRI signal of an elementary WM fiber segment, or {\it fascicle}, can be modeled by two non-exchanging anisotropic Gaussian compartments \citep{Novikov2018b}: an intra-axonal compartment in which axons are represented by impermeable cylinders with zero radius \citep{Veraart2019}; and an extra-axonal compartment represented by an axially symmetric tensor aligned with the fascicle. Given sufficient information in the dMRI measurements, a third, isotropic compartment could be considered. A macroscopic MRI voxel usually contains multiple orientationally dispersed fiber fascicles that contribute to the directional dMRI signal according to their orientation distribution. This general picture, illustrated in Fig.~\ref{fig:overview}d and mathematically defined in {\it Methods} section, is known as the \textit{Standard Model} \citep{Novikov2018b} (SM) of diffusion in WM, as it unifies many previously proposed biophysical models relying on similar assumptions \citep{Kroenke2004,Jespersen2007,Fieremans2011,Zhang2012,Jelescu2015,Kaden2016,Reisert2017,Jespersen2018,Novikov2018,Veraart2018,Lampinen2020,Coelho2022}. 

In this work, we perform a comprehensive histological validation of the SM parameters, by characterizing brain WM microstructure on sham-operated and traumatic brain injury (TBI) rats, using volume (3d) electron microscopy (EM) and \textit{ex vivo} dMRI. The 3d nature of our histological analysis, the variety of the EM samples (corpus callosum, cingulum, in the ipsi- and contralateral sides, for TBI and sham-operated animals), and the large number of segmented 3d axons (in the order of ten thousand per sample) --- all together enable an unprecedented characterization of tissue microstructure. For the dMRI analysis, we estimate the SM parameters using four publicly available approaches: white matter tract integrity (WMTI) \citep{Fieremans2011}, neurite orientation dispersion and density imaging (NODDI) \citep{Zhang2012}, spherical mean technique (SMT) \citep{Kaden2016}, and standard model imaging (SMI) \citep{Reisert2017,Novikov2018,Coelho2022}. We refer to these as {\it SM estimators}, as they target the same biophysical SM parameters, yet differ in the additional constraints (see Table~\ref{tab1:smmethods}) and ways of performing model fitting (max-likelihood versus machine learning). While all these estimators provide SM parameters that correlate significantly ($p<0.05$) with their histological counterparts, indicating sensitivity, we find that SMI shows the highest specificity by presenting smallest cross-correlations with other, non-corresponding histological features. This comparison of SM parameters against 3d EM validates SM metrics as specific tissue biomarkers valuable for clinical applications and neuroscience research, as well as reveals the strengths and limitations of the chosen estimator.

The parameters provided by SM describe specific cellular features and could be a powerful tool for the study of pathological conditions. The \textit{intra-axonal water fraction} $f$ is a potential marker for axon density that could quantify axonal loss \citep{Jespersen2010,VestergaardPoulsen2011,Jelescu2016,CoronadoLeija2021a}, the \textit{intra-axonal diffusivity} $D_\mathrm{a}$ is a potential marker for axon morphology that could quantify axonal injury such as beading \citep{Budde2010,Hui2012,Lee2020}, the \textit{extra-axonal axial diffusivity} $D_\mathrm{e}^\parallel$ is a potential marker for changes in the extra-axonal space such as inflammation, and the \textit{extra-axonal radial diffusivity} $D_\mathrm{e}^\perp$ is a potential marker for demyelination \citep{Fieremans2012,Jelescu2016,Ginsburger2018}. 

Aside from the SM kernel parameters related to properties of  fiber fascicles, SM includes the \textit{fiber orientation distribution} $\mathcal{P}(\hat{n})$ (FOD), which can be described using its spherical harmonics decomposition without assuming any particular functional form. $\mathcal{P}(\hat{n})$ properties can be described using FOD rotational invariants $p_l$, which are the energies of the FOD spherical harmonics coefficients $p_{lm}$ in each degree-$l$ sector (Eq.~\ref{eq:FODpl} in {\it Methods} section). Here, the second-order rotational invariant $p_2$ is of special importance as it is inversely related\citep{Novikov2018}  to the average angle of dispersion of the axons $\theta$ via Eq.~\ref{eq:p2disp1}. $\mathcal{P}(\hat{n})$ and $\theta_{p_2}$ are informative for structural brain connectivity, surgical planning, as well as to assess pruning in neurodevelopment \citep{Jelescu2015}. 

The ability to accurately recover all SM parameters would reveal the wealth of WM tissue microstructure properties encoded in a relatively low-resolution \textit{in vivo} dMRI data. Large initiatives are currently underway to collect extensive \textit{in vivo} dMRI human brain data sets such as the Human Connectome Project \citep{Glasser2016}, the UK Biobank \citep{Miller2016}, the Alzheimer's Disease Neuroimaging Initiative \citep{Jack2008} and the Adolescent Brain Cognitive Development \citep{Casey2018}. Biophysical modeling combined with wide-scale \textit{in vivo} dMRI measurements provides an unprecedented opportunity to study human brain changes in development \citep{Jelescu2015,Mah2017,Lynch2020,Liao2023}, aging \citep{Cox2016,Benitez2018,Beck2021}, and disease \citep{Hui2012,Wang2019,Fieremans2013,Benitez2014,Fu2019,deKouchkovsky2016,Granberg2017,Johnson2021,Palacios2020,Muller2021,Liao2023}. 

\section{METHODS}
\label{sc:methods}

All animal procedures were approved by the Animal Care and Use Committee of the Provincial Government of Southern Finland and performed according to the guidelines set by the European Community Council Directive 86/609/EEC. 

\subsection{Traumatic brain injury model}

Five adult male Sprague-Dawley rats (10 weeks old, weights between 320 and 380 g, Harlan Netherlands B.V., Horst, Netherlands) were used in the study. The animals were housed in a room (22 $\pm$ 1$^\circ$C, 50–60\% humidity) with 12 h light/dark cycle and free access to food and water. 

TBI was induced by lateral fluid percussion injury in three rats \citep{2006Kharatishvili}. Rats were anesthetized with a single intraperitoneal injection, and a craniotomy (5 mm diameter) was performed between bregma and lambda on the left convexity (anterior edge 2.0 mm posterior to bregma; lateral edge adjacent to the left lateral ridge). Lateral fluid percussion injury was induced by a transient fluid pulse impact (21--23 ms) against the exposed intact dura using a fluid-percussion device. The impact pressure was adjusted to 3.2--3.4 atm to induce a severe injury. The sham operation on the other two rats included all the surgical procedures except the impact. See \citep{Salo2018} and \citep{Abdollahzadeh2019} for detailed information about the animal model.

Five months after TBI or sham operation, the rats were transcardially perfused using 0.9\% NaCl for 2 min followed by 4\% paraformaldehyde (PFA). The brains were removed from the skull, post-fixed in 4\% PFA / 1\% glutaraldehyde overnight at 4$^o$C, and then placed in 0.9\% NaCl for at least 12 h to remove excess PFA. 

\subsection{Ex vivo dMRI imaging}

The extracted rat brains were scanned ex vivo at room temperature (21$^o$C) in a vertical 9.4 T/89 mm magnet (Oxford Instruments PLC, Abingdon, UK) interfaced with a DirectDrive console (Varian Inc., Palo Alto, CA, USA) using a quadrature volume RF- coil ({\O} = 20 mm; Rapid Biomedical GmbH, Rimpar, Germany) as both transmitter and receiver. During imaging, the brains were immersed in perfluoropolyether (Solexis Galden \textregistered, Solvay, Houston, TX, USA) to avoid signals from the surrounding area. 

The dMRI data was acquired using a 3d segmented spin-echo EPI sequence with four segments and 0.150$\times$0.150$\times$0.150 mm$^3$ resolution (data matrix 128$\times$96$\times$96, FOV 19.2$\times$14.4$\times$14.4 mm$^3$). TE/TR = 35/1000 ms. The acquisition comprised a total of 132 volumes that consisted in 3 sets of 42 uniformly distributed directions with diffusion weighting (b-values) of 2, 3, and 4 ms/$\upmu$m$^2$, diffusion times $\delta$/$\Delta$ = 6/11.5 ms and three non-diffusion-weighted images. 

\subsection{3d EM imaging}

After the ex vivo dMRI acquisitions, brains were placed in 0.9\% NaCl for at least 4 h to remove excess perfluoropolyether and then sectioned into 1 mm thick coronal sections with a vibrating blade microtome. For each brain, sections at 3.80 from bregma were selected and further dissected into smaller samples containing the areas of interest (Figure~\ref{fig:overview}a and Fig.~\ref{fig:overview}b). Ten WM samples, two from each brain were collected: the ipsi- and contralateral of the cingulum and corpus callosum.  
The samples were stained using an enhanced protocol with heavy metals, dehydrated and embedded in Durcupan ACM resin. Before mounting the specimens, the excess resin in the hardened tissue blocks was trimmed. After selecting the area of interest for imaging within the samples, the blocks were further trimmed into a pyramidal shape with a $1 \times 1$ mm$^2$ base and an approximately $600 \times 600$ $\upmu$m$^2$ top (face), which assured the stability of the block while being cut in the microscope. The tissue was exposed on all four sides, bottom, and top of the pyramid. Silver paint was used to electrically ground the exposed block edges to the aluminum pins, except for the block face or the edges of the embedded tissue. The entire surface of the specimen was then sputtered with a thin layer of platinum coating to improve conductivity and reduce charging during the sectioning process.

The blocks were imaged in a scanning electron microscope (Quanta 250 Field Emission Gun; FEI Co., Hillsboro, OR, USA), equipped with the 3View system (Gatan Inc., Pleasanton, CA, USA) using a backscattered electron detector (Gatan Inc.). The face of the blocks was in the $x$-$y$ plane, and the cutting was in $z$ direction. All blocks were imaged using a beam voltage of 2.5 kV and a pressure of 0.15 Torr. After imaging, Microscopy Image Browser (MIB; \url{http://mib.helsinki.fi}) was used to process and align the EM image stacks. 

Two datasets were acquired simultaneously at low- and high-resolution consistently at one specific location in the white matter in both the ipsi- and contralateral hemispheres for all sham-operated and TBI animals as shown in Fig.~\ref{fig:overview}b. The low-resolution datasets were imaged from large tissue volumes of $200\times 100 \times 60$ $\upmu$m$^3$ with resolution $50\times 50 \times 50$ nm$^3$. Two-thirds of the acquired volumes correspond to the corpus callosum and one-third to the cingulum. Only the corpus callosum region in the ipsilateral side for one rat is missing. The high-resolution datasets were imaged from small tissue volumes of $15\times 15 \times 15$ $\upmu$m$^3$ on the corpus callosum with resolution $15\times 15 \times 50$ nm$^3$. For detailed information about 3d EM tissue preparation, acquisition, and pre-processing, see \citep{Salo2018} and \citep{Abdollahzadeh2019}.    

The labor-intensive sample preparation for EM studies limits the number of samples that can be collected, especially for the large 3d EM samples used in this work \citep{Salo2018}. Yet, thanks to the features of these large 3d EM samples (comparable to the MRI voxel size) and the automated axon segmentation \citep{Abdollahzadeh2019,Abdollahzadeh2021}, this study was able to include a large number of segmented 3d axons per population (in the order of ten thousand) in a diverse set of ROIs, for sham and TBI conditions, allowing for unprecedented microstructure characterization in comparison with previous histological studies.

\subsection{3d EM Segmentation}

Myelin, myelinated axons, and cell nuclei were automatically segmented from the low-resolution datasets using the deep-learning-based pipeline DeepACSON \citep{Abdollahzadeh2021, Abdollahzadeh2021b} (\url{https://github.com/aAbdz/DeepACSON}), which is a method suited for the challenging task of segmenting large datasets with limited visualization of the cell membranes caused by low resolution. 

First, the high-resolution datasets (small tissue volumes) were segmented using the ACSON pipeline \citep{Abdollahzadeh2019}, which is an automated method that integrates edge detection and seeded region growing algorithms in 3d, refining the segmentation with the SLIC (Simple Linear Iterative Clustering) supervoxel technique. These high-resolution images and their corresponding segmentations were down-sampled to match the low-resolution images (large tissue volumes) and used as human-annotation-free training sets for the DeepACSON pipeline \citep{Abdollahzadeh2021}. DeepACSON uses two convolutional neural networks to compute probability maps of myelin, myelinated axons, mitochondria, cell nuclei, and cell nuclei membranes. Then, these probability maps are binarized, and the segmentations are refined. Because of the low resolution of the images, the myelinated axons are prone to under-segmentation; for this reason, the segmented axons are rectified using a cylindrical shape decomposition method \citep{Abdollahzadeh2021b}, and finally a support vector machine is used to exclude non-axonal segmented objects. 

Further proofreading was performed by considering only the myelin attached to the segmented axons, thus removing spurious segmented objects in the extra-axonal space that were labeled as myelin. For the computation of volume fractions all segmented axons were considered, but for the quantifications of properties such as: axon diameter, cross-sectional area variation, and the fiber orientation distribution; only axons larger than 10 $\upmu$m, were used. In order to avoid biases due to oblique cross sections at the extremes caused by the inclination of the axons, 1 $\upmu$m was removed at the two ends of each axon.  

\subsection{Microstructural metrics derived from 3d EM}

Using the segmented 3d EM samples, the intra-axonal, extra-axonal, and myelin volumes were computed as the number of voxels covering each compartment. The dMRI related intra-axonal volume fraction metric $f$ was computed as the intra-axonal volume divided by the total volume of the sample, excluding the myelin compartment due to its short $T_2$ compared with the TE of the dMRI acquisition \citep{1994MacKay}, which makes it dMRI invisible. 

Morphological properties of the segmented axons were obtained following the approach in \citep{Lee2019} (\url{https://github.com/NYU-DiffusionMRI/RaW-seg}). To compute axon diameters and cross-sectional area variations, each axon's main direction was first aligned to the $z$-axis. Then, the axon skeleton was created as the line connecting the center of mass of each slice. Assuming a circular cylinder axon, the cross-sectional area $A(z)$ and the diameter $2r\equiv2\sqrt{A(z)/\pi}$ of a circle with the same area were computed for each slice perpendicular to the axon skeleton. 

Recently, the exact axial tortuosity ($\Lambda_{\parallel}$) relation between the varying cross-sectional areas $A(z)$ along the main axis of the axon and the intra-axonal diffusivity was derived \citep{Abdollahzadeh2023} from the Fick-Jacob’s equation:
\begin{equation}
\Lambda_{\parallel}=\frac{D_0}{D_{\mathrm{a}}} = \left\langle\frac{\overline{A}}{A(z)}\right\rangle,
\label{eq:AxialTort}
\end{equation}
where $\overline{A}$ is the mean cross-sectional area, and $D_0$ is the diffusivity for a perfect cylinder axon.   

Dispersion angle for an axon population is defined \citep{Novikov2018} as: 
\begin{equation}
\theta=\cos^{-1}\sqrt{ \langle \cos^2 \theta_i \rangle },
\label{eq:DispAng}
\end{equation}
where the individual axon segment’s dispersion angle $\theta_i$ is the angle between the axon segment’s direction with respect to the main direction of the axon population. Dispersion angle $\theta_u$ caused only by undulations \citep{Lee2020} can be computed using Eq.~\ref{eq:DispAng} on the axons aligned to the z-axis. 

Assuming free diffusivity for perfect cylinders ($D_0=D_\mathrm{w}=2.0\ \upmu\mathrm{m}^2/\mathrm{ms}$ for dMRI acquired at room temperature), the value of SM intra-axonal diffusivity $D_a$ can be predicted from histology using Eq.~\ref{eq:AxialTort}, and $\theta_u$ computed with Eq.~\ref{eq:DispAng} on each axon, aggregating the contribution for all of them: 
\begin{equation}
\tilde{D}_\mathrm{a}=\sum_{k} w_{k}\frac{D_\mathrm{w}}{\Lambda_{k}^{\parallel}}\cdot \langle \cos^2 \theta_u \rangle_{k},
\label{eq:PredictedDa}
\end{equation}
where the weights $w_{k}$ are proportional to the axon volumes, with $\sum_{k}w_{k}\equiv1$.

Fiber orientation distributions on each WM region for each animal were computed using the corresponding skeletons of the aligned axons by rotating them back to the original main directions and projecting their tangent vectors on a 3d triangulated spherical surface \citep{Lee2019}; creating in this way a histogram $\mathcal{P}(\hat{n})$ that was normalized such that $\int\! \mathrm{d}\hat{n} \mathcal{P}(\hat{n})\equiv 1$. These FODs were decomposed on spherical harmonic (SH) basis:
\begin{equation}
\mathcal{P}(\hat{n}) \approx 1+\sum_{l=2,4,...}^{l_{max}}\sum_{m=-l}^{l}p_{lm}Y_{lm}(\hat{n}),
\label{eq:FODSH}
\end{equation}
where $Y_{lm}$ are the SH basis, and $p_{lm}$ are the SH coefficients. Because of FOD antipodal symmetry, only even harmonic orders $l$ are considered up to a maximum order of $l_{max}=16$. Also, for a normalized FOD $p_0=1$. To simulate coarse-grained effects caused by diffusion in the calculation of the FOD, the axon skeletons were smoothed by a Gaussian kernel with variance $\sigma^2=L^2/4$ \citep{Novikov2018b,Lee2019}, where $L=\sqrt{2Dt}$ is the diffusion length, $D=2\upmu\mathrm{m}^2/\mathrm{ms}$, and $t$ is the diffusion time of the dMRI acquisition.

From the SH decomposition of $\mathcal{P}(\hat{n})$ in Eq.~\ref{eq:FODSH}, FODs rotational invariants $p_l$ were be computed \citep{Reisert2017,Novikov2018}:
\begin{equation}
p_{l}^2=\frac{1}{4\pi(2l+1)}\sum_{m=-l}^{m=l}|p_{lm}|^2, 
\label{eq:FODpl}
\end{equation}
where $p_2$ is of special interest as it is related with the average dispersion angle of the axons \citep{Novikov2018,Lee2019} in Eq.~\ref{eq:DispAng} by
\begin{equation}
\theta\approx\theta_{p_2}\equiv \cos^{-1}\sqrt{\frac{2p_2+1}{3}}.
\label{eq:p2disp1}
\end{equation}
Some models represent the FOD $\mathcal{P}(\hat{n})$ as a Watson distribution \citep{Zhang2012,Jespersen2018}, which is defined by its concentration parameter $\kappa$. A relation between $\kappa$ and $p_2$ is given in \citep{Jespersen2018}:
\begin{equation*}
p_2 = \frac{1}{4}\left( \frac{3}{\sqrt{\kappa}F(\sqrt{\kappa})} - 2 - \frac{3}{\kappa} \right).
\label{eq:p2kappa}
\end{equation*}

It has been suggested \citep{Reisert2017} that the rotational invariants $p_l$ of the FODs follow an exponential decay with respect to the harmonic order $l$, given by the Poisson kernel $ p_l=p_0 \cdot \lambda^l$ with $p_0 = 1$ and $\lambda\leq1$. In \citep{Lee2019}, by using segmented 3d axons in the normal mouse corpus callosum, it was found that this functional form was not properly normalized and an extra term was required
\begin{equation}
p_l=C \lambda^l,
\label{eq:pldecay}
\end{equation}
where $C\geq 1$, for $l=2,4,...$. In our work, given the large number of segmented 3d axons, and the variety of samples with and without pathology, this hypothesis was further validated.

\begin{table*}[htb!]
\begin{center}
\begin{tabular}{lccc} \\ \hline 
\textbf{Method} & \textbf{Independent} & \textbf{Parameter} & \textbf{FOD} \\ 
& \textbf{parameters}& \textbf{constraints} & \\ \hline
 & & & \\
\textbf{WMTI} \citep{Fieremans2011}  & $f,D_\mathrm{a},D_\mathrm{e}^\parallel,D_\mathrm{e}^\perp$ & $D_\mathrm{a}\leq D_\mathrm{e}^\parallel$ & Coherently aligned fibers \\
 & & & \\
  & & $D_\mathrm{a}=D_\mathrm{e}^{\parallel}=0.6$ $\upmu$m$^2$/ms & \\
\textbf{NODDI} \citep{Zhang2012} & $f,f_\mathrm{w},\kappa$ & $D_\mathrm{e}^{\perp}=D_\mathrm{e}^{\parallel}\cdot(1-f)$ & Watson distribution \\
  & & & \\
 & & $D^{\parallel}=D_\mathrm{a}=D_\mathrm{e}^{\parallel}$ & \\
\textbf{SMT} \citep{Kaden2016}   & $f,D^{\parallel}$ & $D_\mathrm{e}^{\perp}=D^{\parallel}\cdot(1-f)$ & Factored out by spherical mean \\
 & & & \\
\textbf{SMI} \citep{Reisert2017,Coelho2022}   & $f,D_\mathrm{a},D_\mathrm{e}^\parallel,D_\mathrm{e}^\perp,f_\mathrm{w},p_{lm}$ & Unconstrained & Unconstrained \\ 
& & & \\ \hline
\end{tabular}
\caption{\textbf{Standard Model parameter estimators}. Although all these approaches aim to estimate essentially the same SM parameters, the main differences between them are the number of independent parameters (second column), the constraints/interdependencies imposed on parameters (third column), the functional form of the FOD (fourth column), as well as different strategies for estimation (WMTI: linear diffusion and kurtosis tensors fit followed by explicit analytical formulas based on kurtosis tensor parameters; NODDI and SMT: max-likelihood; SMI: machine learning approach based on the polynomial regression from signal's rotational invariants to SM paramteres). All these differences lead to different values in parametric maps, Fig.~\ref{fig:maps}, and correlations with histological counterparts, Figs.~\ref{fig:correlations}--\ref{fig:scatterplots}.}
\label{tab1:smmethods}
\end{center}
\end{table*}

\subsection{Standard Model}

The Standard Model \citep{Novikov2018b} of the dMRI signal in the measured direction $\hat{u}$ for brain WM (Figure~\ref{fig:overview}d), can be written as:
\begin{equation}
S(b,\xi,\hat{u}) = S_0\int_{|\hat{n}|=1} \mathrm{d}\hat{n} \mathcal{P}(\hat{n}) \mathcal{K}(b,\xi,\hat{u}\cdot \hat{n}).
\label{eq:SMConv}
\end{equation}
This equation describes the dMRI signal $S(b,\xi,\hat{u})$ as a convolution between the FOD $\mathcal{P}(\hat{n})$
and the SM kernel $\mathcal{K}(b,\xi,\hat{u}\cdot \hat{n})$, which represents the fiber segment, and is the sum of two, non-exchanging, intra- and extra-axonal Gaussian compartments: 
\begin{equation*}
\mathcal{K}(b,\xi,\hat{u}\cdot \hat{n})=fe^{-bD_\mathrm{a}(\hat{u}\cdot \hat{n})^2}+(1-f)e^{-bD_\mathrm{e}^\perp-b(D_\mathrm{e}^\parallel-D_\mathrm{e}^\perp)(\hat{u}\cdot \hat{n})^2},
\label{eq:SMKernel}
\end{equation*}
where $b$ is the diffusion weighting, $\hat{u}\cdot \hat{n}$ is the cosine of the angle between the symmetry axes of the SM kernel $\hat{n}$ and the measured direction $\hat{u}$, and $\xi$ are the parameters of the SM kernel:
\begin{equation}
\xi=\left\{ f,D_\mathrm{a},D_\mathrm{e}^\parallel,D_\mathrm{e}^\perp \right\}.
\label{eq:smpars}
\end{equation}

In SH basis, the convolution between the FOD and the SM kernel in Eq.~\ref{eq:SMConv} is a product
\begin{equation}
s_{lm}(b,\xi)=p_{lm}\cdot \mathcal{K}_l(b,\xi).
\label{eq:SMSH}
\end{equation}
Orientation information can be factored out by computing the signal, FOD and SM kernel rotational invariants \citep{Reisert2017,Novikov2018}:
\begin{equation}
s_{l}(b,\xi)=p_{l}\cdot \mathcal{K}_l(b,\xi),
\label{eq:SMSHrotinv}
\end{equation}
with
\begin{equation}
s_{l}^2=\frac{1}{4\pi(2l+1)}\sum_{m=-l}^{m=l}|s_{lm}|^2, 
\label{eq:rotinvS}
\end{equation}
for the diffusion MRI signal, and
\begin{equation}
\mathcal{K}_l(b,\xi)=\int_{0}^{1}\mathrm{d}x \mathcal{K}(b,\xi,x)P_l(x), 
\label{eq:kl}
\end{equation}
for the SM kernel, where $P_l$ are the Legendre polynomials. 

The estimation of the SM parameters from conventional Pulsed-Gradient Spin-Echo (PGSE) dMRI is challenging because the problem is ill-posed \citep{Jelescu2015b,Novikov2018}. It has been shown that ``advanced" dMRI acquisitions are needed for an accurate, precise, and robust estimation of all the SM metrics \citep{Novikov2018,Lampinen2020,Coelho2022}. Nonetheless, there is still a need to correctly estimate the SM parameters on the typical PGSE data used in this study, as the results can be extrapolated to the majority of clinical human studies that use similar dMRI protocols including the large multicenter dMRI datasets currently being collected \citep{Glasser2016,Miller2016,Jack2008,Casey2018}. Several strategies have been proposed to solve the degeneracy of the SM parameter estimation in PGSE data. The principal characteristics of these strategies, regarding to estimated parameters and constraints, are summarized in Table~\ref{tab1:smmethods}.

\textit{White matter tract integrity} \citep{Fieremans2011} uses the diffusion and kurtosis tensors to derive formulas for the SM parameters, however, its mostly valid for aligned axon populations. \textit{Neurite orientation dispersion and density imaging} (NODDI) \citep{Zhang2012} is the most widely used estimator for $f$ and the FOD (as Watson distribution), however, it fixes the axial diffusivities, and constrains the extra-axonal radial diffusivity using the first order tortuosity approximation \citep{Szafer1995}. \textit{Spherical mean technique} (SMT) \citep{Kaden2016} uses the spherical mean (0$^{th}$ order rotational invariant) on each shell of the dMRI signal to factor out the FOD and focus only in the estimation of the SM kernel parameters, however, it uses two of NODDIs strong assumptions. \textit{Standard model imaging} (SMI) \citep{Reisert2017,Coelho2022} uses a machine learning approach to solve the maximum-likelihood estimation of the system \citep{Novikov2018} in Eq.~\ref{eq:SMSHrotinv}, here, the rotational invariants of the dMRI signal (Eq.~\ref{eq:rotinvS}) are mapped to the SM kernel parameters (Eq.~\ref{eq:smpars}) via cubic polynomial regression, however, when the information in the data is insufficient, the estimation of the parameters could be biased by the training data.

SM parameters also include the FOD $\mathcal{P}(\hat{n})$, which can be described without assuming any functional form. Once the SM kernel parameters $\xi$ are obtained, the FODs SH coefficients $p_{lm}$ can be estimated from Eq.~\ref{eq:SMSH}. Also, the FODs SH rotational invariants $p_{l}$ can be directly estimated from Eq.~\ref{eq:SMSHrotinv}. 

\begin{figure*}[htb!] 
\centering
\includegraphics[width=0.99\textwidth]{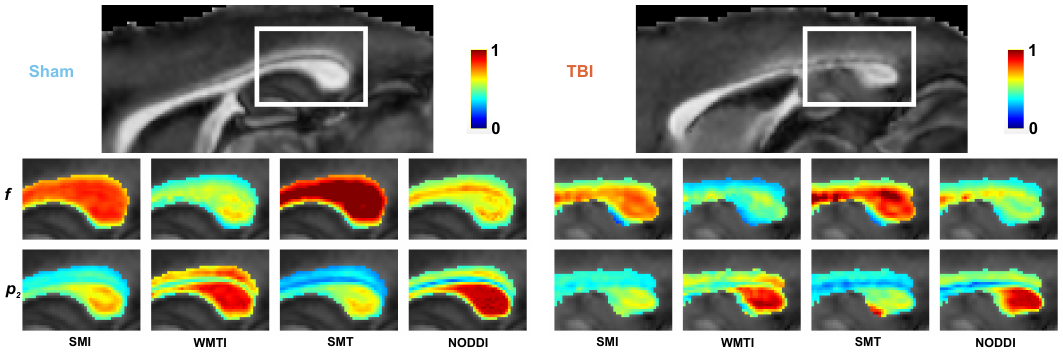}
\caption{\textbf{Standard Model parameter maps from different estimators}. Maps for $f$ (intra-axonal water fraction) and $p_2$ (inversely related to dispersion) from SMI, NODDI, WMTI, and SMT for a sham-operated and a TBI animal. Changes in TBI with respect to sham, such as the reduction in $f$ and $p_2$, agree with histology for all methods. However, evident quantitative differences between their maps are observed.}
\label{fig:maps} 
\end{figure*}

\subsection{dMRI Processing}

In order to improve the precision and accuracy of the estimated SM maps, dMRI data was processed following the steps in the DESIGNER pipeline (\url{https://github.com/NYU-DiffusionMRI/DESIGNER}) \citep{AdesAron2018} adapted for \textit{ex vivo} dMRI rat brains. Magnitude and phase volumes were reconstructed from the data to perform MP-PCA complex-denoising \citep{Veraart2016,LEMBERSKIY2020}, which has shown to highly reduce the Rician noise floor. Images were also corrected for Gibbs-ringing \citep{Kellner2015}, and motion distortions \citep{AVANTS2008}. WMTI maps were computed using tools provided by the DESIGNER pipeline. NODDI, SMT and SMI maps were estimated with their respective toolboxes: SMT (\url{https://github.com/ekaden/smt}), NODDI (\url{http://mig.cs.ucl.ac.uk/index.php?n=Tutorial.NODDImatlab}), and SMI (\url{https://github.com/NYU-DiffusionMRI/SMI}). 

For NODDI, the axial diffusivity was fixed to the suggested value for \textit{ex vivo}, $D^{\parallel}=D_\mathrm{a}=D_\mathrm{e}^{\parallel}=0.6\ \upmu\mathrm{m}^2/\mathrm{ms}$. For SMT and SMI, the upper bound for the diffusivities was set to the free diffusion of water at room temperature $\sim 2\ \upmu\mathrm{m}^2/\mathrm{ms}$. For SMI, rotational invariants up to $l=6$ were used for parameter mapping. SM \textit{free water fraction} $f_\mathrm{w}$ metric can be estimated by SMI and NODDI, however, imaging was performed using $b = 2,3,4 $ ms/$\upmu$m$^2$ at room temperature (21 $^o$C), for which free water signal is negligible ($\approx 1.5\%$ of $S_0$ for the lower b-vale). For this reason, for SMI we set $f_\mathrm{w}=0$. However, for NODDI $f_\mathrm{w}$ was estimated as it was shown to improve the correlations of its other parameters. Manually delineated ROIs were drawn by identifying, in the diffusion tensor ellipsoids, the corpus callosum, cingulum and their crossing region in the ipsi- and contralateral side at the location where the 3d EM samples were collected (Fig.~\ref{fig:S00}).

FODs second-order rotational invariant $p_2$ is sensitive to axon dispersion, as shown in Fig.~\ref{fig:dispersion}a, and its affected by complex fiber configuration such as crossing fibers in the voxel. In the dMRI volumes, we observed partial volume in the cingulum ROI which contained a small fraction of corpus callosum axons in the perpendicular direction. For this reason, the SM maps in Fig.~\ref{fig:overview}f and Fig.~\ref{fig:maps} show increased dispersion (higher $\theta_{p_2}$ and lower $p_2$) in the cingulum voxels with respect to the corpus callosum voxels, contrary to what it is observed in the 3d EM volumes. SMI estimates the full non-parametric FOD, from which we segmented the lobes corresponding to the different axon populations \citep{Riffert2014}, then from the segmented lobes we computed per-bundle rotational invariants $p_l$. 

\subsection{Statistics}

Pearson correlations ($\rho$), p-values ($p$), and 95\% confidence intervals ($CI$) between each dMRI and 3d EM derived metric pairs were computed using MATLABs command \textbf{corrcoef}, which uses a t-test with $N-2$ degrees of freedom for p-values ($N$ is the number of samples), and for the confidence intervals uses an asymptotic Gaussian distribution of $\frac{1}{2}\ln\left(\frac{1+\rho}{1-\rho}\right)$, with an approximate variance equal to $1/(N-3)$. Here, the total number of samples used in the analysis is 28 instead of 30, as the corpus callosum region in the 3d EM is missing, hence no cc and no cc+cg, for one sham-operated animal. We corrected for multiple comparisons using the False Discovery Rate (FDR) computed with MATLABs command \textbf{mafdr} that uses the Benjamini–Hochberg procedure. Concordance correlation coefficients $\rho_c$ were also computed between SM dispersion angle $\theta_{p_2}$ and its histological counterpart $\theta$, as agreement in their values is expected for these metrics in particular. It is important to mention that in this study a correlation does not necessarily imply causality, as a pathological condition such as TBI is very complex, with many microstructural changes happening simultaneously, and it is challenging to get the full picture of how a particular SM parameter is affected by them. Also, statistical significance, in our case $p<0.05$, could be affected by the number of samples and the assumptions made by the statistical test, such as Gaussianity. 

\begin{figure*}[htb!]
    \centering
    \includegraphics[width=0.99\textwidth]{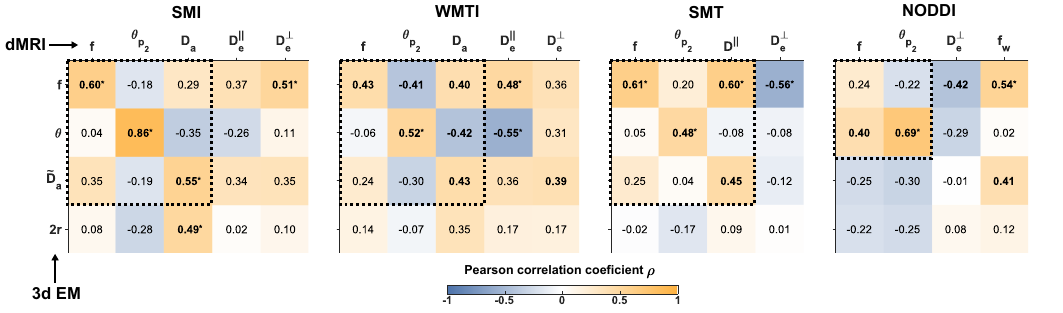}
    \caption{\textbf{Comparison between dMRI and 3d EM derived metrics}. For each estimator, matrices show the Pearson correlation coefficients comparing 3d EM histology derived metrics (rows) against dMRI estimated SM parameters (columns), as plotted in Fig.~\ref{fig:scatterplots}. Significant correlations ($p<0.05$) are highlighted in bold, where * indicates that significance remains after adjusting for multiple comparisons using the false discovery rate. Supplementary Figs.~\ref{fig:S01}-\ref{fig:S04} lists also the p-values and 95\% confidence intervals. Submatrices for corresponding dMRI and 3d EM metrics, in which significant correlations are expected only in the diagonal, are highlighted by dashed lines. }
    \label{fig:correlations}
\end{figure*}

\section{RESULTS}
\label{sc:results}

The TBI rat brains show visible damage due to the impact on 1 mm-thick brain sections (Fig.~\ref{fig:overview}a) and corresponding fractional anisotropy (FA) maps (Fig.~\ref{fig:overview}c), as opposed to sham-operated rat brains showing no visible damage. To reveal the corresponding changes caused by TBI at the microstructural scale, we automatically segmented the myelin and myelinated axons on 3d EM, as illustrated in Fig.~\ref{fig:overview}b. Fig.~\ref{fig:overview}e shows the reduction of axonal density, changes in axon morphology (diameter $2r$, axial tortuosity $\Lambda_{\parallel}$ and undulations, as defined in {\it Methods} section), as well as an increase in axonal dispersion in the TBI sample. Remarkably, these microstructural changes are also detected non-invasively by ex vivo MRI, as shown on the SM maps in Fig.~\ref{fig:overview}f for the axon water fraction $f$, intra-axonal diffusivity $D_\mathrm{a}$, and axon dispersion angle $\theta_{p_2}$. 

\begin{figure*}[htb!]
    \centering
    \includegraphics[width=0.99\textwidth]{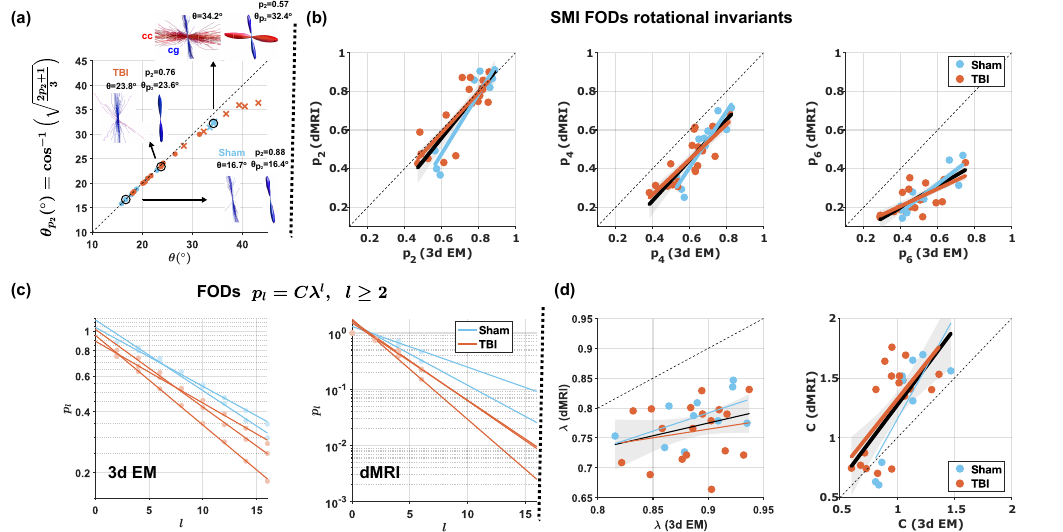}
    \caption{\textbf{Dispersion angle $\theta$, FODs rotational invariants $p_l$, and their exponential decay}. (a) For 3d EM histology, dispersion angle $\theta_{p_2}$ (computed from $p_2$) corresponds to the true average angle of deviation $\theta$ of the axons with respect to the main direction of the axon population. The agreement is better for single bundle FODs ($\cdot$ markers) than for crossing fibers (x markers). (b) FODs rotational invariants $p_2$, $p_4$, $p_6$ correlate with their histological counterparts. (c) FODs $p_l$ follow an exponential decay $p_l=C\lambda^l$ for $l=2,4,...$. Examples from dMRI and 3d EM in the ipsilateral cingulum for all animals. (d) Comparison between dMRI (SMI) and 3d EM derived parameters $C$ and $\lambda$.}
    \label{fig:dispersion}
\end{figure*}

Parametric maps obtained from the same ex vivo dMRI measurement but using the different estimators (SMI, WMTI, NODDI, SMT) are compared in Fig.~\ref{fig:maps} for $f$ and $p_2$ (its relation to dispersion angle via Eq.~\ref{eq:p2disp1} is demonstrated in Fig.~\ref{fig:dispersion}a). Consistent changes are observed between sham-operated and TBI animals that are in qualitative agreement with the histology shown in Fig.~\ref{fig:overview}e. Yet, the maps also demonstrate quantitative differences among the distinct estimators.

Sensitivity and specificity are evaluated by comparing SM parameters $f$, $\theta_{p_2}$ and $D_\mathrm{a}$ with their 3d EM counterparts on six ROIs: cc, cg, cc+cg for the ipsi- and contralateral hemispheres for each animal, as shown in Fig.~\ref{fig:S00}. While histological $f$ and $\theta$ are computed directly from the segmented axons, $\tilde{D}_\mathrm{a}$ is predicted from histology (Eq.~\ref{eq:PredictedDa}) by considering the axons axial tortuosity \citep{Abdollahzadeh2023} $\Lambda_{\parallel}$ and the dispersion angle caused only by axon undulations \citep{Lee2020} $\theta_u$, assuming free diffusivity $D_\mathrm{w}= 2\ \upmu\mathrm{m}^2/\mathrm{ms}$ (at room temperature) for perfect cylinder axons. The influence of axon diameter $2r$ in the SM parameters is also evaluated.    

The diagonals of the dashed submatrices in  Fig.~\ref{fig:correlations} indicate that all SM estimators provide metrics that correlate significantly ($p<0.05$) with their corresponding tissue properties derived from 3d EM histology, except for $f$ from NODDI. This suggests good overall sensitivity of all SM estimators. Here, SMI achieves the highest correlations for the three SM metrics, with only SMT providing a similar correlation for $f$.  

Regarding specificity, the off-diagonal elements of the dashed submatrices in Fig.~\ref{fig:correlations} show that only SMI has no significant correlations with non-corresponding histological features, while the other estimators reveal multiple spurious cross-correlations. Looking at each SM parameter individually (columns), dMRI $f$ is specific to its histological counterpart for SMI, WMTI, and SMT, while for NODDI, it correlates with 3d EM $\theta$. dMRI $\theta_{p_2}$ is specific to histological $\theta$ for SMI, SMT and NODDI, while for WMTI, it correlates also with 3d EM $f$. dMRI $D_\mathrm{a}$ is specific to its predicted value from histology $\tilde{D}_\mathrm{a}$ only for SMI, while for SMT, it correlates also with 3d EM $f$, and for WMTI, it shows correlations also with 3d EM $f$ and $\theta$. The highest specificity of SMI parameters can be understood by the lack of hard constraints and assumptions as opposed to the other estimators (see Table~\ref{tab1:smmethods}).

\begin{figure*}[htb!]
    \centering
    \includegraphics[width=1.02\textwidth]{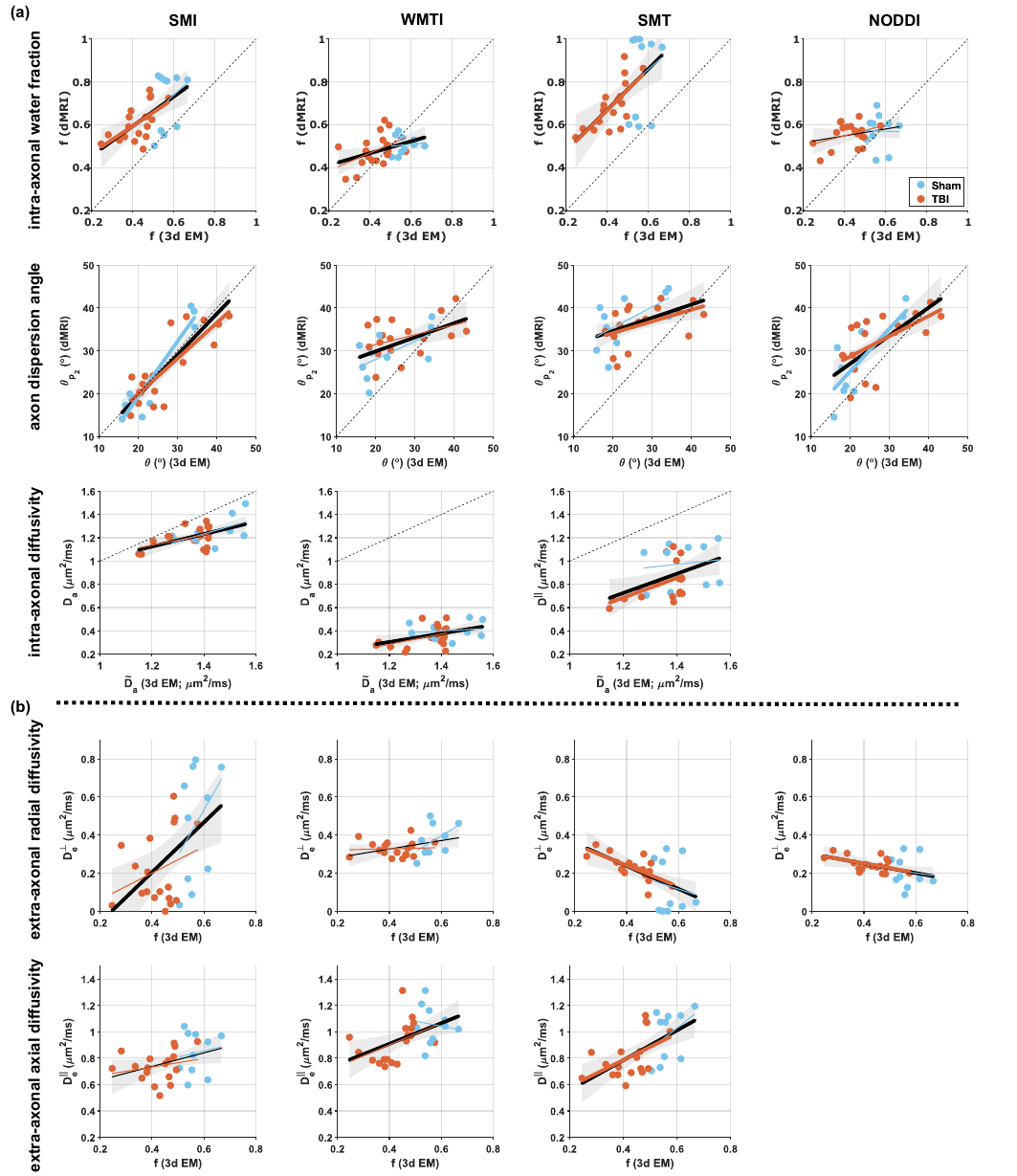}
    \caption{\textbf{Scatter plots for dMRI derived parameters against histology metrics}. Linear regressions are shown using all (black lines and their 95\% confidence intervals), TBI-only (red lines) and sham-only (blue lines) samples. Thicker lines indicate a correlation with $p<0.05$. The numerical values (using all samples) for the Pearson correlations coefficients are shown in Fig.~\ref{fig:correlations}, their p-values and 95\% confidence intervals are listed in Fig.~\ref{fig:S01}-\ref{fig:S04}. (a) SM parameters $f$, $D_\mathrm{a}$, and $\theta_{p_2}$ from SMI, WMTI, SMT, and NODDI compared against their corresponding 3d EM metrics. (b) SM extra-axonal diffusivities $D_\mathrm{e}^{\perp}$ and $D_\mathrm{e}^{\parallel}$ compared against histological $f$.}
    \label{fig:scatterplots}
\end{figure*}

In the scatter plots for dMRI $f$ against EM $f$ of Fig.~\ref{fig:scatterplots}a, agreement with the identity line is not expected. For one part, the resolution of the EM samples does not allow to segment unmyelinated axons. Furthermore, dMRI data was acquired using a single TE, so SM $f$ is $T_2$-weighted, while EM $f$ is not. For these reasons, and assuming intra-axonal $T_2$ longer than extra-axonal $T_2$, \citep{Veraart2018,Tax2021} an overestimation of dMRI $f$ is expected. 

On the other hand, agreement with the identity line is expected when comparing dMRI $\theta_{p_2}$ against EM $\theta$ in the scatter plots of Fig.~\ref{fig:scatterplots}, as unmyelinated axons are expected to have similar dispersion than myelinated axons, and the fiber orientation distribution is expected to be unaffected by relaxation. In Fig.~\ref{fig:scatterplots}a, better agreement with the identity line is observed for $\theta_{p_2}$ from SMI than from the other methods, this is also indicated by its highest Pearson correlation ($\rho$) values shown in Fig.~\ref{fig:correlations} and its highest concordance correlation coefficients $\rho_c$, with respect to the other methods: SMI $\rho_c=0.850$, WMTI $\rho_c=0.332$, SMT $\rho_c=0.191$, and NODDI $\rho_c=0.569$.

In the scatter plots of Fig.~\ref{fig:scatterplots}a that compare dMRI $D_\mathrm{a}$ against its predicted value from histology $\tilde{D}_\mathrm{a}$ (Eq.~\ref{eq:PredictedDa}), we observe that $D_\mathrm{a}$ from SMI agrees better with the identity line. For the other methods, the estimation of $D_\mathrm{a}$ could be biased because of SMT constraint $D^{\parallel}=D_\mathrm{a}=D_\mathrm{e}^{\parallel}$, and WMTI assumption $D_\mathrm{a}\leq D_\mathrm{e}^\parallel$. NODDI keeps $D_\mathrm{a}$ fixed and therefore the scatter plots for this metric are not shown. Correlation between $D_\mathrm{a}$ and axon diameter is observed for SMI, but not for the other methods, however, it presents the lower value among the significant correlations of SMI.    

Figure~\ref{fig:scatterplots}b scatter plots reveal positive correlations for SMI and WMTI $D_\mathrm{e}^\perp$ with histological $f$, while negative correlations are observed for SMT and NODDI. The negative correlations follow from the extra-axonal tortuosity approximation used by NODDI and SMT \citep{Szafer1995,Zhang2012,Kaden2016}, that relates $D_\mathrm{e}^\perp$ with $f$ and $D_\mathrm{e}^\parallel$ as shown in Table~\ref{tab1:smmethods}, and as also observed in Monte Carlo simulation studies that mimic axonal loss \citep{Szafer1995,Fieremans2012}. SMI and WMTI, on the other hand, don't impose this constraint, and show a somewhat surprising opposite sign of correlation. 

Using the segmented 3d axons, we validate in Fig.~\ref{fig:dispersion}a that the dispersion angle $\theta_{p_2}$, computed from the histological FOD $p_2$ (Eq.~\ref{eq:p2disp1}), is a good approximation for the true angle of dispersion $\theta$ of the axons. While dMRI $p_2$, and hence $\theta_{p_2}$, can be straightforwardly obtained from all SM estimators (Eq.~\ref{eq:SMSHrotinv}), SMI also estimates non-parametric FODs in SH basis. The rotational invariants $p_l$ computed from these dMRI FODs (Eq.~\ref{eq:FODpl}) correlate with their corresponding histological $p_l$, as shown in Fig.~\ref{fig:dispersion}b. These higher order $p_l$ allow to further validate their exponential decay \citep{Reisert2017,Lee2019} with respect to the increasing order $l$, for $l\geq2$ in Eq.~\ref{eq:pldecay}, for both 3d EM and dMRI, as shown in Fig.~\ref{fig:dispersion}c for the ipsilateral cingulum of all animals. Furthermore, by comparing the exponential decay parameters $C$ and $\lambda$, from SMI and histology, we observe in Fig.~\ref{fig:dispersion}d that $C$ shows better agreement. It should be noted that dMRI parameters $C$ and $\lambda$ are directly computed from the rotational invariants $p_2$, $p_4$ and $p_6$ estimated with SMI, without imposing the exponential decay functional form as constraint during the fitting.

\section{DISCUSSION}

Biophysical modeling bridges the resolution gap between the macroscopic dMRI voxel and microscopic tissue features, and provides a powerful tool to  study development, aging, and diseases at the cellular level where these processes originate. Numerous human \textit{in vivo} studies in brain WM \citep{Jelescu2015,Mah2017,Lynch2020,Cox2016,Benitez2018,Beck2021,Hui2012,Wang2019,Fieremans2013,Benitez2014,Fu2019,Palacios2020,Muller2021,deKouchkovsky2016,Granberg2017,Johnson2021,TakeshigeAmano2022,Liao2023} have been conducted using SM parameters derived from different estimators: WMTI, NODDI, SMT and SMI. However, the interpretation of these and future studies significantly relies on the biological accuracy and promised microstructural specificity of the metrics provided by these estimators \citep{Jelescu2017,Novikov2018a}, which highly depends on how correctly they describe the relevant features of the cellular environment, and how accurately and precisely they are estimated.    

Previous validation efforts so far have been limited  to 2d histology or incomplete. Indeed, the intra-axonal water fraction $f$ has been evaluated in rodents by comparing against axon density or axon volume fraction metrics derived from 2d optical and electron microscopy (EM) for normal tissue \citep{Jespersen2010}, and models for traumatic brain injury (TBI) \citep{Wang2013}, demyelination \citep{Jelescu2016}, and axonal degeneration \citep{CoronadoLeija2021a}; $f$ also has been evaluated in postmortem human tissue for multiple sclerosis \citep{Grussu2017}. However, the use of 2d microscopic images limited the scope of these studies, since 3d tissue properties, such as the FOD and axon morphology metrics (that reveal injuries along the axons), could not be quantified, and specificity of $f$ was not evaluated. Other studies have compared the FOD and/or dispersion parameters against related histological metrics derived, not from segmented axons, but from structure tensor analysis of 2d optical microscopy \citep{Leergaard2010,Grussu2017,Mollink2017}, 3d confocal \citep{Schilling2018} and 3d electron microscopy \citep{Salo2021}. However, the previous 3d studies focused on healthy tissue and did not evaluate any other metric besides dispersion, preventing the validation of the specificity of SM metrics. Furthermore, previous histological works did not assess the differences between the available SM estimators, such as WMTI, NODDI, SMT, and SMI, and focused the analysis on only one of them, biasing the results to the particular estimator used in the study. 

In this work, we characterized WM microstructure of sham-operated and TBI rat brains using 3d EM and \textit{ex vivo} dMRI to validate the sensitivity and specificity of the SM parameters. The large number of segmented 3d axons per sample (in the order of the ten thousand), and the variety and large size of these samples (from corpus callosum and cingulum, in the ipsi- and contralateral hemispheres of the brain, for sham-operated and TBI rats), allowed an unprecedented histological microstructure characterization (Figure~\ref{fig:overview}e) revealing the following WM microstructure changes due to TBI: reduction in axon density, injuries along the axon (observed by the reduction in axon diameter, and by the increase in beading and undulation), and increase in axon dispersion. The simultaneous changes in several microstructural properties indicate the complexity of the damage in chronic TBI, and the need for sensitive and specific dMRI biomarkers in order to properly disentangle between them.

The direct comparison of 3d EM derived $f$ vs dMRI intra-axonal water fraction $f$  (Figure~\ref{fig:correlations}) reveals the strongest sensitivity for $f$ estimated with SMT and SMI, followed by WMTI, as well as strong specificity for these estimators, as there are no correlations between $f$ and $\theta$ nor $D_\mathrm{a}$. On the other hand, $f$ from NODDI correlates with $\theta$ but not with histological $f$, which may suggest that imposing constraints biases NODDI parameter estimation, especially in pathological scenarios \citep{Jelescu2020}. Nonetheless, the otherwise overall agreement between SM estimators in terms of sensitivity and specificity is in line with the consistent findings in many human \textit{in vivo} studies that report changes in $f$ with development, aging and disease. Indeed, $f$ has been shown to increase in development and decrease in aging using NODDI \citep{Jelescu2015,Mah2017,Lynch2020,Cox2016,Beck2021,Liao2023}, WMTI \citep{Jelescu2015,Benitez2018,Beck2021,Liao2023}, SMT \citep{Beck2021,Liao2023} and SMI \citep{Liao2023}. In neurological disorders, $f$ is observed to increase in stroke ischemic lesions using NODDI \citep{Wang2019}, WMTI \citep{Hui2012}, and SMI \citep{Liao2023}, while decrease in TBI \citep{Palacios2020,Muller2021} using NODDI, Alzheimer's disease using WMTI \citep{Fieremans2013,Benitez2014} and NODDI \citep{Fu2019}, and in Parkinson's disease \citep{TakeshigeAmano2022} using NODDI. Furthermore, $f$ has been suggesting as surrogate marker to track disease as it correlates with disability scores in multiple sclerosis, using WMTI \citep{deKouchkovsky2016}, NODDI, SMT \citep{Johnson2021}, and SMI \citep{Liao2023}. The consistent correspondence between $f$ estimated by WMTI, SMT, and SMI, with 3d EM $f$ shows that changes in this metric can be interpreted as mainly driven by changes in axon volume fractions. 

For dispersion, Fig.~\ref{fig:correlations} shows that the dMRI dispersion angle $\theta_{p_2}$ is sensitive and specific to the 3d EM true angle of dispersion of the axons $\theta$ for SMI, SMT, and NODDI, while for WMTI it shows to be sensitive but not specific, correlating also with histology $f$. This histological validation may help interpret human \textit{in vivo} studies that showed changes in dMRI derived axon dispersion, namely reduced dispersion in Alzheimer's disease \citep{Fu2019}, and chronic TBI \citep{Muller2021}, while increased dispersion in stroke \citep{Wang2019,Liao2023}, multiple sclerosis \citep{Granberg2017,Liao2023}, and Parkinson's disease, \citep{TakeshigeAmano2022}. The quantitative differences obtained by distinct SM estimators, as shown in Fig.~\ref{fig:maps} and Fig.~\ref{fig:scatterplots}a, are also observed in human \textit{in vivo} studies using different estimators. In particular, dispersion has shown to decrease during neurodevelopment using WMTI \citep{Jelescu2015} and SMI \citep{Liao2023} in the genu corpus callosum, which has been associated with axonal pruning, while weak or no significant changes have been observed using NODDI \citep{Jelescu2015,Mah2017,Lynch2020,Liao2023}. Our histological validation helps understand these discrepancies and highlight the importance of SM estimators and their underlying constraints. Interestingly, SMI provides a good one-to-one correspondence with histological dispersion angle $\theta_{p_2}$, which confirms that the lack of hard constraints (Table~\ref{tab1:smmethods}) in SM estimation results in the best overall accuracy for SMI \citep{Liao2023}. 

Using microscopy data to validate compartmental specific diffusivities $D_\mathrm{a}$, $D_\mathrm{e}^\parallel$, and $D_\mathrm{e}^\perp$ is challenging, as there is no straightforward way to directly match them with histological metrics \citep{Jelescu2017}. However, insight can be gained from studying changes in these diffusivities under certain pathological conditions \citep{Jelescu2016,Budde2010,Hui2012,Lee2020}, or by performing Monte Carlo simulations \citep{Budde2010,Lee2020,Abdollahzadeh2023} in realistic substrates mimicking brain tissue in different (pathological) states. Here, we compared SM $D_\mathrm{a}$ against the expected diffusivity from histology $\tilde{D}_\mathrm{a}$ (Figure~\ref{fig:scatterplots}a and Eq.~\ref{eq:PredictedDa}) based on the axon tortuosity $\Lambda_{\parallel}$ \citep{Abdollahzadeh2023} and the dispersion from axon undulation $\theta_u$ \citep{Lee2020}. Correlations between $D_\mathrm{a}$ and $\tilde{D}_\mathrm{a}$ are observed for SMI, SMT and WMTI. However, specificity is observed only for $D_a$ from SMI, as $D_a$ from SMT and WMTI both correlates with other non-corresponding histological features. Our result underscore the previously observed dependency of $D_\mathrm{a}$ on axon diameter variation and axon beading, obtained using Monte Carlo simulations \citep{Budde2010,Lee2020,Abdollahzadeh2023} and human \textit{in vivo} studies \citep{Hui2012,Lee2020,Liao2023}. 

For the extra-axonal perpendicular diffusivity $D_\mathrm{e}^\perp$, Fig.~\ref{fig:scatterplots}b shows different dependencies on EM-derived volume fraction between the different SM estimators. The positive correlation between histological $f$ and SMI/WMTI $D_\mathrm{e}^\perp$, contradicts the extra-axonal tortuosity approximation used as a hard constraint in NODDI and SMT. While intuitively, $D_\mathrm{e}^\perp$ is expected to decrease with increasing $f$, Monte Carlo simulations of axon loss have also shown opposite trends \citep{Szafer1995,Fieremans2012}. The behavior of $D_\mathrm{e}^\parallel$ and $D_\mathrm{e}^\perp$ during pathological conditions is still an ongoing topic of investigation, as the complex processes happening in the extra-axonal space are not fully understood. Our results suggest that using functional relations between  $D_\mathrm{e}^\perp$ and $f$ is unjustified, and both parameters should rather be estimated independently, particularly in pathological conditions. Interestingly, for SMI and WMTI that do not employ the tortuosity constraint, several human \textit{in vivo} studies have found changes in their $D_\mathrm{e}^\perp$ for neurodevelopment \citep{Jelescu2015,Liao2023}, aging \citep{Benitez2018,Beck2021}, stroke \citep{Hui2012,Liao2023}, Alzheimer's disease \citep{Fieremans2013,Benitez2014}, and multiple sclerosis \citep{deKouchkovsky2016,Liao2023}.

As an outlook, this work bridges the gap between micrometer-scale cellular architecture and the macroscopic MRI resolution, by comprehensively validating a key white matter diffusion MRI model and its four publicly available estimators, WMTI, NODDI, SMT, and SMI. The uniqueness of the present approach is in the 3-dimensional EM, the use of pathology (TBI) as well as sham-operated animals, the variety of the samples, and the large number of segmented 3d axons. The specific correspondence with histology indicates that the validated open-source Standard Model estimator will become a powerful tool for neuroscience research, providing valuable non-invasive imaging markers to study brain development, aging, disease, disorders and injuries at the cellular level where these processes originate.

\section{Data and code availability}

The 3d EM volumes and their segmentation are publicly available at \url{https://etsin.fairdata.fi/dataset/f8ccc23a-1f1a-4c98-86b7-b63652a809c3}. The dMRI volumes will also be made publicly available. Several publicly available toolboxes were used in this study as indicated in the methods section. Extra in-house codes created for the analysis of the 3d EM and dMRI data will be made publicly available by the authors.

\section{Funding}

Research was supported by the National Institute of Neurological Disorders and Stroke of the NIH under awards R01  NS088040 and R21  NS081230, and by the Hirschl foundation, and was performed at the Center of Advanced Imaging Innovation and Research (CAI2R, www.cai2r.net), a Biomedical Technology Resource Center supported by NIBIB with the award P41  EB017183. A.S. was funded by the Academy of Finland (grant \#323385) and the Erkko Foundation.  H.H.L. was funded by the Office of the Director of the NIH under award DP5 OD031854.

\section{Author contributions} 
R.C.L, E.F., and D.S.N. conceived the project and designed the study. R.C.L. performed the formal analysis of the 3d EM and dMRI data, extracted the microstructural metrics and compared them. A.A. segmented the 3d EM data. A.S. provided animal models, dMRI and EM imaging. H.H.L., S.C., B.A.A., and Y.L. provided software for the processing and analysis of the EM and dMRI data. A.A., J.T, and A.S. contributed with the processing and segmentation of the 3d EM data. R.A.S. contributed with the analysis of the dMRI data. R.C.L. and E.F. wrote the manuscript. E.F. and D.S.N. supervised the project. All authors commented, edited and approved the final manuscript.

\section{Competing interests}
E.F. and D.S.N. are co-inventors in technology related to this research with US patents US10360472B2 and US10698065B2.

\newpage
\newpage

\bibliography{bibliography}


\clearpage
\newpage
\begin{titlepage}

\begin{center}
\Large\textbf{Supplementary Information}\\[1cm]
\Large{Volume electron microscopy in injured rat brain validates white matter microstructure metrics from diffusion MRI}\\[0.5cm]
\large{Ricardo Coronado-Leija$^{1,*}$, Ali Abdollahzadeh$^1$, Hong-Hsi Lee$^2$, Santiago Coelho$^1$, Benjamin Ades-Aron$^1$, Raimo A. Salo$^3$, Jussi Tohka$^3$, Alejandra Sierra$^3$, Dmitry S. Novikov$^1$, Els Fieremans$^1$} \\[2cm]
\end{center}

\noindent $^1$Bernard and Irene Schwartz Center for Biomedical Imaging, Department of Radiology, New York University School of Medicine, New York, NY, USA

\noindent $^2$Athinoula A. Martinos Center for Biomedical Imaging, Department of Radiology, Massachusetts General Hospital, Harvard Medical School, Boston, MA, USA

\noindent $^3$A.I. Virtanen Institute for Molecular Sciences, University of Eastern Finland, Kuopio, Finland

\noindent $^*$ricardo.coronadoleija@nyulangone.org, rleija@cimat.mx
\end{titlepage}

\pagenumbering{roman}
\setcounter{page}{1}


\renewcommand{\thepage}{S\arabic{page}} 
\renewcommand{\thefigure}{S\arabic{figure}} 
\setcounter{figure}{0} 
\setcounter{table}{0}

\begin{figure*}[htb!]
    \centering
    \includegraphics[width=0.99\textwidth]{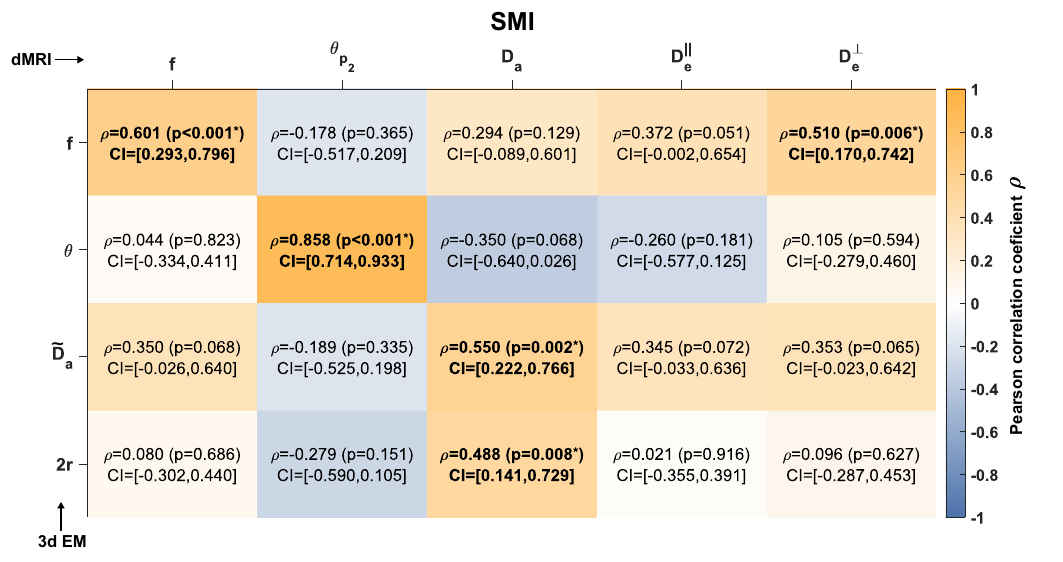}
    \caption{\textbf{Comparison between SMI parameters and 3d EM derived metrics}. Each cell shows the Pearson correlations coefficient, p-values and 95\% confidence intervals. Significant correlations ($p<0.05$) are highlighted in bold, where * indicates that significance remains after adjusting for multiple comparisons using the false discovery rate.}
    \label{fig:S01}
\end{figure*}

\begin{figure*}[htb!]
    \centering
    \includegraphics[width=0.99\textwidth]{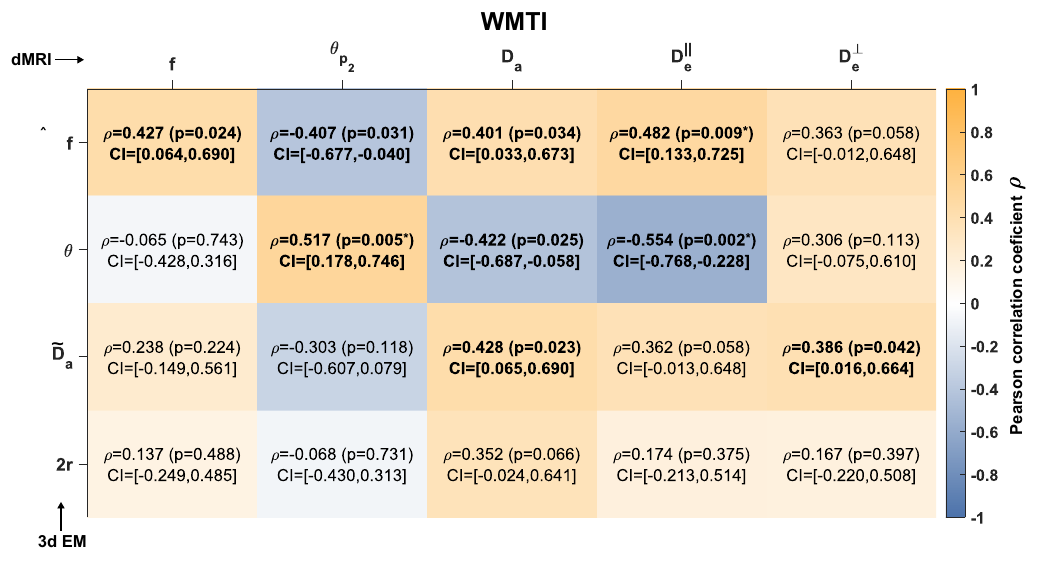}
    \caption{\textbf{Comparison between WMTI parameters and 3d EM derived metrics}. Each cell shows the Pearson correlations coefficient, p-values and 95\% confidence intervals. Significant correlations ($p<0.05$) are highlighted in bold, where * indicates that significance remains after adjusting for multiple comparisons using the false discovery rate.}
    \label{fig:S02}
\end{figure*}

\begin{figure*}[htb!]
    \centering
    \includegraphics[width=0.99\textwidth]{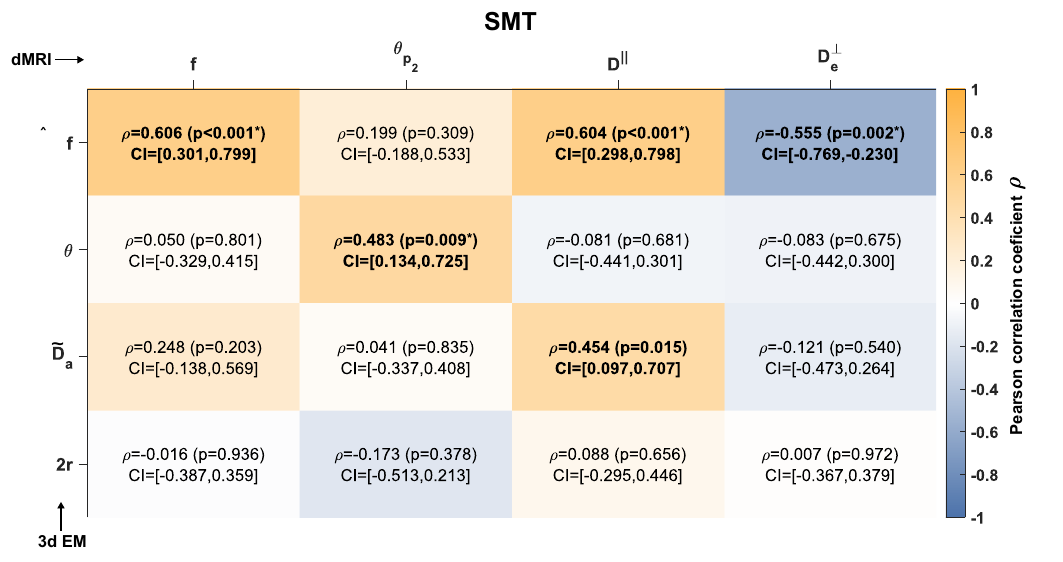}
    \caption{\textbf{Comparison between SMT parameters and 3d EM derived metrics}. Each cell shows the Pearson correlations coefficient, p-values and 95\% confidence intervals. Significant correlations ($p<0.05$) are highlighted in bold, where * indicates that significance remains after adjusting for multiple comparisons using the false discovery rate.}
    \label{fig:S03}
\end{figure*}

\begin{figure*}[htb!]
    \centering
    \includegraphics[width=0.99\textwidth]{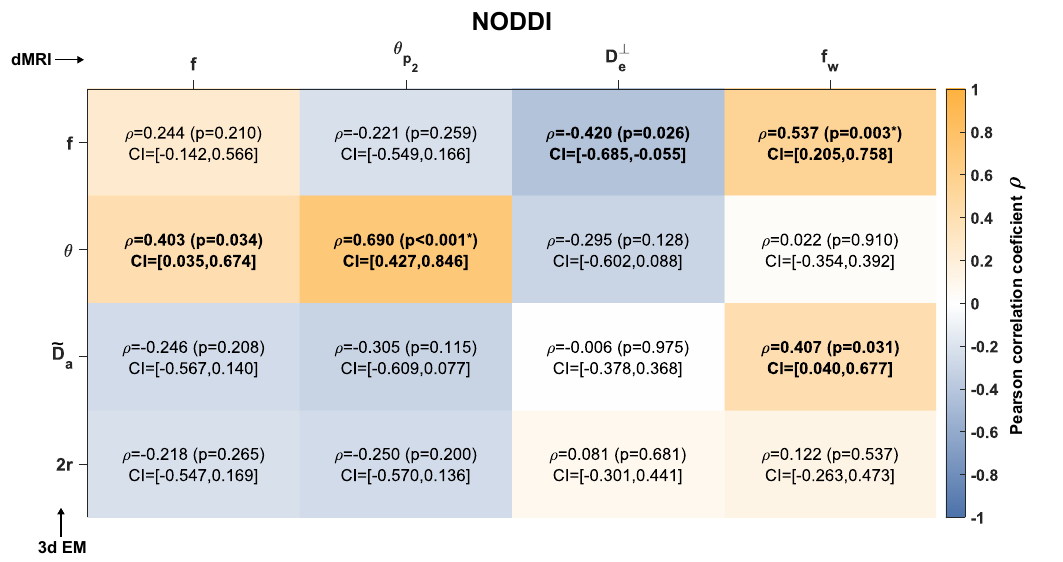}
    \caption{\textbf{Comparison between NODDI parameters and 3d EM derived metrics}. Each cell shows the Pearson correlations coefficient, p-values and 95\% confidence intervals. Significant correlations ($p<0.05$) are highlighted in bold, where * indicates that significance remains after adjusting for multiple comparisons using the false discovery rate.}
    \label{fig:S04}
\end{figure*}

\begin{figure*}[htb!]
    \centering
    \includegraphics[width=0.99\textwidth]{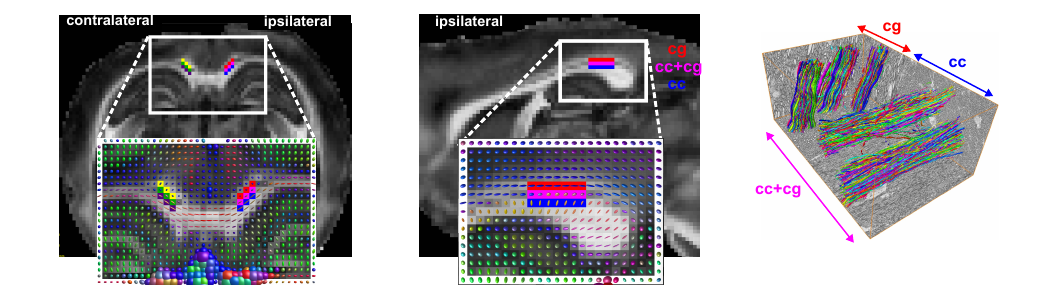}
    \caption{Six ROIs were analyzed on each animal: cc, cg and cc+cg for the ipsi- and contralateral hemispheres. }
    \label{fig:S00}
\end{figure*}

\end{document}